\documentclass[12pt,showkeys]{revtex4}
\begin{document}

\title{A Note on the Integral Formulation of Einstein's Equations
Induced on a Braneworld}
\author{Christine C. Dantas} 
\email{ccdantas@iae.cta.br}
\altaffiliation{Instituto de Aeron\'autica e Espa\c co (IAE), 
Centro T\'ecnico Aeroespacial (CTA), P\c ca. Mal. Eduardo Gomes, 50,
CEP 12.228-904 - Vila das Ac\'acias
S\~ao Jos\'e dos Campos - SP - Brazil}
\date{\today}

\begin{abstract}

We revisit the integral formulation (or Green's function approach) of
Einstein's equations in the context of braneworlds. The integral
formulation has been proposed independently by several authors in the
past,  based on the assumption that it is possible to give a
reinterpretation of the local metric field in curved spacetimes as an
integral expression involving sources and boundary conditions.  This
allows one to separate source-generated  and source-free contributions
to the metric field. As a consequence, an exact meaning to Mach's
Principle can be achieved in the sense that only source-generated
(matter fields) contributions to the metric are allowed for; universes
which do not obey this condition would be non-Machian. In this paper,
we revisit this idea concentrating on a Randall-Sundrum-type model
with  a non-trivial cosmology on the brane.  We argue that the role of
the surface term (the source-free contribution) in the braneworld
scenario may be quite subtler than in the 4D formulation. This may
pose, for instance, an  interesting issue to the cosmological constant
problem.

\end{abstract}







\maketitle

Mach's Principle (MP) is often understood as the general idea that
inertia arises from the interaction of matter with the rest of all
matter in the universe. It is well known that this idea played a
fundamental  role in the developments of general relativity (GR). In
1917, Einstein \cite{ein1917} added the so-called cosmological
constant, $\Lambda$, to his field equations in order to find a static
solution to his cosmological model, in which a direct connection of
the mass density in the universe to its geometry could be achieved in
accordance with Mach's ideas.  However, there are solutions to
Einstein's field equations that simply fail to  be Machian. For
instance, as early as in 1917, de Sitter \cite{des} proposed a static
cosmological model including $\Lambda$ but no matter fields as a
solution to Einstein's equations.  Since then,  the validity of
MP remains an open question of GR \cite{mp}. There are in fact dozens
of possible formal interpretations for this principle in the context
of different classical theories of gravitation (see the index on page
530 of Ref. \cite{mp} for the various possible definitions of MP).  At
the same time, the so-called `cosmological constant problem' is  one
of the most important unsolved issues in physics today (see reviews
in, e.g.,  \cite{wein}, \cite{pad}). Therefore it is tempting to
inquire whether MP and the cosmological constant could be deeply
related in some manner \cite{vish}.

On the other hand, it is important to realize that the question of the
origin of inertia as well as the existence of a non-vanishing
cosmological constant, as indicated by present-day cosmological
observations \cite{knop}, \cite{ton}, \cite{sper},  are not only a
problem of classical theories but also a fundamental issue that any
consistent quantum theory of gravity must address. Indeed, the
construction of a consistent quantum theory of the gravitational field
is the main goal of current theoretical physics. Presently, the two
most relevant approaches are Loop Quantum Gravity \cite{rovelli} and
String/M-Theory (see reviews and introductory lectures in \cite{m}).
The former focuses on the search for a relational notion of a quantum
spacetime, whereas the latter is more appropriately regarded as an
ambitious research program towards finding an unified,
non-perturbative description of all fundamental interactions of nature
\cite{smolin}.  Although the literature on these new theories is vast,
there has been very few discussions regarding the problem of the
origin of inertia as advocated by MP in high energy physics (some work
can be found in Ref. \cite{mygh}; interpretations of MP in the context
of quantum gravity can be found in Ref. \cite{mp}, Chap. 8).

In most quantum gravity theories, GR is supposed to break down at high
energies (the Planck scale, at $\sim 10^{19}$ GeV). In some
string-inspired models (`braneworld scenarios'), gravity emanates as a
higher-dimensional theory and 4-dimensional Einstein's GR is expected
to be reproduced at low enough energies as an effective theory (e.g.,
\cite{branes}). These braneworld models state in general that our
observable universe (viz. all Standard Model fields) is confined to a
`3-brane', i.e., a (3+1)-dimensional subspace of a higher,
five-dimensional (`bulk') space-time, where the gravitational field
extends along a fifth-dimension. In the  Randall-Sundrum type 2 model
(RS2) \cite{rs2} model, for instance, there is only one three-brane
and a large extra-dimension. In this case, the general relativistic
gravity is well reproduced in the low energy limit \footnote{The
Randall-Sundrum type 1 model  (RS1) \cite{rs1}, on the other hand,
attempts to solve the hierarchy problem by a small extra dimension,
with two three-branes, one with negative and  the other with positive
tension. This mechanism solves the hierarchy problem only if we live
on the negative tension brane, but in this case gravity is not
localized on this brane.}.

In the present letter, we briefly discuss such a model in the context of  the
integral formulation  of GR, generalized here to five dimensions.
The original formalism (see, e.g., \cite{inte}, \cite{mwt}) was
developed independently by Sciama, Waylen \& Gilman \cite{swg},
Al'tshuler \cite{al} and Lynden-Bell \cite{lynd}.  The idea is that,
despite the nonlinearity of Einstein's field equations,  a
reinterpretation of the metric potentials in curved space-time as an
integral expression involving sources and boundary conditions can be
given.    To that end, the information locked into the gravitational
field of distant matter is propagated linearly through a
self-consistent curved space-time (viz. the space-time derived from
the sum of the contributions of all the matter in the universe).  For
instance, suppose that the 4D metric tensor $g_{\mu \nu}$, which
ultimately determines the local inertial frames of reference, is
completely derived from a joint influence of the matter-energy content
in the universe, then the metric could be expressed as:
\begin{equation}
g_{\mu \nu}(x^{\mu}) = 
2  \int_{\Omega}
{
G_{\mu \nu \beta^{\prime}}^{~~~~\alpha^{\prime}}
(x^{\mu},x^{{\mu}^{\prime}})
K_{\alpha^{\prime}}^{~~\beta^{\prime}}(x^{\mu^{\prime}})
[-g(x)]^{1/2} d^4x^{\prime}
} + \int_{\partial \Omega} 
{\nabla_{\gamma\prime}}G_{\mu \nu \beta^{\prime}}^{~~~~\alpha^{\prime}} 
g_{~~\alpha^{\prime}}^{~\beta^{\prime}}(x^{{\mu}^{\prime}})
[-g(x)]^{1/2} dS^{\gamma^{\prime}}.
\end{equation}
The source function is given by 
\begin{equation}
K_{\alpha^{\prime}}^{~~\beta^{\prime}}(x^{\mu^{\prime}}) \equiv 
\kappa^2 \left [ 
T_{\alpha^{\prime}}^{~~\beta^{\prime}}(x^{\mu^{\prime}})
- 1/2T_{\gamma^{\prime}}^{~~\gamma^{\prime}} (x^{\mu^{\prime}})
\delta_{\alpha^{\prime}}^{~~\beta^{\prime}} \right ]
- \Lambda
\delta_{\alpha^{\prime}}^{~~\beta^{\prime}},
\end{equation}
with $\kappa^2 \equiv 8 \pi G$, and 
$\Lambda$ is the cosmological constant. 
The source function is therefore related to the energy-momentum tensor
for the cosmic fluid and possible vacuum energy density contributions
are allowed.  The Green function, $G_{\mu
\nu}^{\alpha^{\prime}\beta^{\prime}}(x^{\mu},x^{{\mu}^{\prime}})$ is a
second rank tensor at the two space-time points $x^{\mu}$ and
$x^{{\mu}^{\prime}}$, namely, a propagator operator of the
gravitational field of a body located at $x^{{\mu}^{\prime}}$ to
coordinate $x^{\mu}$.   The second integral of Eq. (1) is a surface
term representing the contribution to the metric from the data
specified on the intersection of the observer's light cone
$\Gamma_x^{-}$ and the surface $\delta \Omega$ bounding the proper
volume $\Omega  \equiv \sqrt{(-g)} d^4x^{\prime}$  of spacetime.   The
symbol $\nabla_{\gamma}$ denotes the covariant derivative,  and
$dS^{\gamma^{\prime}}$ is the coordinate surface element on $\delta
\Omega$, which points outward from $\Omega$. Notice that in the
classical GR, the Green function must sharply go to zero outside the
past light cone of an observer at $x$, and therefore the first
integration term must proceed over the inner region of space delimited
by $\Gamma_x^{-}$ and $\delta \Omega$.  Notice also that the metric in
Eq. (1) is not itself a solution to Einstein's equations, but an
equivalent representation of them as integral equations.

The integral formalism has the advantage of formally separating
source-generated and source-free contributions to the metric
field. Gilman \cite{gil} provides a classification scheme for
interpreting cosmological models under MP in a strict sense, that is,
the joint contribution of all mass-energy  of matter in the
universe fully specifies the metric tensor $g_{\mu \nu}$, up to
diffeomorphisms. Notice that gravitational (geometrical) degrees of
freedom do not enter in the above definition of MP.  In these terms, a
cosmological model is considered Machian from Gilman's scheme if it
does not contain source-free contributions [the surface term of
Eq. (1)  vanishes, the so-called `Gilman condition'].  In fact, the
surface integral is to be considered as an integral representation of
the solution to the homogeneous wave equation, namely, Einstein's
field equations with a null source term. In globally hyperbolic
spacetimes, the integral formulation as a whole is a well-defined
representation for both the homogeneous as well as the nonhomogeneous
field equations.  The boundary surface may be at an infinite proper
past, or may well be formed by the the union of particle horizons for
all points along the observer's timelike worldline.  When the volume
of spacetime contains all the sources in the observer's past light
cone, then the surface term is to be interpreted as contributions to
the local metric field that come from matter outside the volume of
integration and/or from the data at the boundary surface, or, in
the words of Gilman,  as `contributions to the local field that cannot
be attributed to any of the observable sources' \cite{gil}.  Notice
that such boundary surface must be an {\it initial} surface where
conditions must be specified. For instance, if we impose the following
constraint equations on the initial surface,
$\nabla_{\gamma^{\prime}}G_{\mu \nu \beta^{\prime}}^{~~~~\alpha^{\prime}} =0,
$ then these constraints are equivalent to the Machian requirement that
the surface term, which includes source-free contributions to the
local metric field, initially vanishes.

Gilman further assumes that, in a Machian cosmology, $\Lambda$
must be identically zero,  since for the integral formulation to be
valid (Eq. 1), $\Lambda$ must be   treated as a source term
\cite{swg}, as is generically assumed in Eq. (2) above, but according
to the strict definition of MP, it is not a Machian term (it unrelated
to matter fields).  Hence, the `Gilman condition' for a strict Machian
cosmology assumes that the surface term and the cosmological constant
in Eq. (1)  must be zero. In the present paper, we will relax the
latter definition  of MP above, in the sense that the
energy-momentum tensor of the vacuum, $\langle T_{\mu \nu}^{vac}
\rangle$, can be considered as a source term that contributes to the
specification of the metric tensor as well.  Explicitly, we assume
that (e.g., \cite{wein}): $\langle T_{\mu \nu}^{vac} \rangle = 
- \langle \rho^{vac} \rangle g_{\mu \nu}$,
where $\langle \rho^{vac} \rangle$ is the energy density of 
the vacuum (the 4D metric tensor $g_{\mu \nu}$ 
has signature $-+++$ in our notation), in the four-dimensional universe.
Such definition leaves the form of the classical Einstein's equations,
\begin{equation}
G_{\mu \nu} \equiv R_{\mu \nu} - {1 \over 2}R g_{\mu \nu}= - \Lambda 
g_{\mu \nu} + \kappa^2 T_{\mu \nu},
\end{equation}
unaltered if the cosmological constant is redefined as:
$\Lambda \rightarrow \Lambda + \kappa^2 
\langle \rho^{vac} \rangle$.
Such a setup is just one of the `many faces' of the cosmological
constant (see \cite{pad}).
 
The Einstein's equations on the 3-brane world resulting from 
the Randall-Sundrum type 2 model \cite{rs2}, generalized to
allow for matter fields on
the brane were deduced by Shiromizu et al. \cite{branes},
where  higher-dimensional modifications to the standard
Einstein's equations were explicitly found. The induced
field equations on the brane are:
\begin{equation}
\hat{G}_{\mu \nu} = -\Lambda_{\rm eff}\hat{g}_{\mu \nu}
+\kappa^2 \hat{T}_{\mu \nu},
\end{equation}
where
\begin{equation}
\hat{T}_{\mu \nu} \equiv
T_{\mu \nu}
 + {[\kappa^{(5)}]^4 \over \kappa^2 } \pi_{\mu \nu} -
{1 \over \kappa^2} E_{\mu \nu}
\end{equation}
is a redefined energy-momentum tensor, and the tensors $\pi_{\mu \nu}$
and $E_{\mu \nu}$ are correction terms that arise in the induced
Einstein's equations due to higher-dimensional effects. We see that
these terms can all be absorbed into the redefined energy-momentum
tensor, ultimately leaving the form of the induced Einstein's
equations on the brane unchanged from their standard (4D)
expressions.   The correspondences between the 4D and 5D quantities
(scales) involved are the following: first, the induced cosmological
constant on the brane is 
\begin{equation}
 \Lambda_{\rm eff} = {1 \over 2}
[\kappa^{(5)}]^2 \left (\Lambda^{(5)} + {1 \over 6} [\kappa^{(5)}]^2
\lambda^2 \right ),
\end{equation}
and $\lambda = 6 {\kappa^2\over
[\kappa^{(5)}]^4}$ is the so-called brane tension. Notice that
$\Lambda_{\rm eff}$ receives contributions from the bulk
cosmological constant, $\Lambda^{(5)}$, and the brane tension, so that
both could be fine-tuned in order to give a null cosmological constant
on the brane \footnote{Notice that in the original RS2 model, the
cosmological constant on the brane is exactly zero.}.  For low enough
energy densities (namely, lower than the brane tension), gravity is
effectively 4D for the brane observer.  Also, one important point is
that Newton's gravitational constant, $G$, depends on the brane
tension.  Second, the tensor $\pi_{\mu \nu}$ gives local corrections,
quadratic in the energy-momentum tensor, due to matter fields on the
brane (see \cite{branes}), and, finally, the influence of the free
gravitational field in the bulk is expressed as the projection of the
5D Weyl tensor onto the brane, namely, $ E_{\mu \nu} =
{~}^{(5)}C^A_{~~BCD}n_{A}n^{C}g_{\mu}^{~~B}g_{\nu}^{~~D}$,  where
$n^{A}$ is the unit vector normal to the brane.

Now we turn our attention to the integral formulation of GR in this
braneworld scenario. It is immediately clear that, because of the fact
that  the form of the induced Einstein's equations remain  unchanged
from their standard (4D) expressions, the same happens to the
corresponding integral-induced expressions.  In other words, a first
trivial interpretation would be to consider  the additional corrections to
the classical field equations, represented by $\pi_{\mu \nu}$ and
$E_{\mu \nu}$,  simply as additional source terms to Eq. (2) in the
case of the braneworld.  The volume of integration would extend to the
whole brane and include the higher-dimensional effects as additional
source terms.  Matter motions excite gravitational waves in the
bulk, and, on the other hand, the excitations of the free
gravitational field also perturb the dynamics of matter on the
brane. These influences are all taken as source terms in the volume
integral. However, what happens to the surface term in this case?
Could it be interpreted in the usual manner, restrited to a surface at
infinite past on the observer's light cone on the brane?

In the usual (4D) interpretation,  when the volume of spacetime
contains all the sources in the observer's past light cone, the
surface term is to be interpreted as contributions to the local metric
field that come from matter outside the volume of integration and/or
from the data at the boundary surface.  But we immediately see that,
in the braneworld scenario,  (i) the surface of integration must
extend to the bulk; or (2) the higher-dimensional degrees of freedom
must be included in the surface term computed on the brane.  Either
alterations represent non-negligible additional effects 
depending on the Machian conditions imposed to the braneworld observer.

In order to address these questions, 
a note of caution is necessary. As already mentioned,
the metric in Eq. (1) does not represent a solution to Einstein's 
field equations, but is only an equivalent representation of them as 
integral equations. Such equations are not supposed to
be solved with any iteration scheme in RS2 cosmologies, because the
induced Einstein's equations derived by Shiromizu et al. are not closed. 
Notice that the tensor $E_{\mu \nu}$ is responsable for transmitting the 
effects of nonlocal gravitational degrees of freedom from the bulk to the
brane. Solutions to the induced Einstein's equations  on the brane
will generally depend on the evolution of the gravitational field in
the bulk as well, in a somewhat complicated manner (see Shiromizu et
al.\cite{branes}). Hence, in order to qualitatively elaborate the 
role of MP in the braneworld scenario from the integral approach,
we are led to assume a decomposition of the $E_{\mu \nu}$ tensor into
a transverse-traceless part, $E^{TT}_{\mu \nu}$ (corresponding to the
free gravitational degrees of freedom in 5D), and into a longitudinal
part, $E^{L}_{\mu \nu}$ (corresponding to matter field contributions
on the brane). Under such assumptions, the induced Einstein's equations 
are completely closed with respect to the brane quantities {\it if} 
$E^{TT}_{\mu \nu}$ were zero \cite{branes}. This property leads us to
conjecture the possibility of splitting the contribution of $E_{\mu
\nu}$ into a part directly related to a source term on the brane 
($E^{L}_{\mu \nu}$), 
and a source-free contribution ($E^{TT}_{\mu \nu}$) to the the surface term.

A simple example elaborates on further possibilities. We separate the 
contributions of the effective cosmological constant (Eq. 6) into a bulk, 
source-free contribution, to be included in the surface term, 
and a source-generated contribution from the tension of the brane,
to be included in the volume integration term. With such a setup,
from Gauss' theorem, it is  possible to impose, for instance, constraints such as:
\begin{equation}
\int_{\partial {\rm brane}} 
{\nabla_{\gamma\prime}}G_{\mu \nu}^{\alpha^{\prime}\beta^{\prime}} 
\hat{g}_{\alpha^{\prime}\beta^{\prime}}(x^{{\mu}^{\prime}})
[-\hat{g}(x)]^{1/2} dS^{\gamma^{\prime}}\equiv
\int_{\rm brane}{
\nabla_{\rho^{\prime}}^2 \mathcal{E}_{\mu \nu}
[-\hat{g}(x)]^{1/2} d^4x^{\prime}},
\end{equation}
with
\begin{equation}
\mathcal{E}_{\mu \nu} \propto
-[\kappa^{(5)}]^2\Lambda^{(5)}
\hat{g}_{\mu \nu}
- E_{\mu \nu}^{TT}.
\end{equation}
Notice that the bulk cosmological constant is not interpreted
in this setup as a source term corresponding to the bulk vacuum
energy density: we have isolated any contributions from the 
vacuum energy density to the brane tension.           
In other words, the bulk cosmological constant
enter here only as a term that modifies
the 5D gravity Lagrangian to
 $L_{\rm grav} \propto R^{(5)} - 2\Lambda^{(5)}$.
We have reinterpreted the surface term as to
include  source-free contributions solely from the bulk via the
$\mathcal{E}_{\mu \nu}$ tensor defined above. 
In order that a Machian braneworld satisfies the `Gilman condition',
the surface term must go to zero as it  tends to the infinite past,
leading to the  following constraint:
$ \nabla_{\rho^{\prime}}^2 \mathcal{E}_{\mu \nu} = 0.$
In particular, if the bulk is
purely AdS ($ E_{\mu \nu}^{TT} = 0$) and Machian,  the bulk cosmological
constant must satisfy $\nabla_{\rho^{\prime}}^2 \Lambda^{(5)} 
{g}_{\mu \nu} = 0$.
In general, such a simple setup indicates that, for a Machian braneworld,
the components of $\mathcal{E}_{\mu \nu}$  cannot increase or
decrease in all directions from a given spacetime location at the
initial hypersurface defined from the observer's infinite past
light-cone. Other considerations can be equally made on such grounds.

In summary,  we believe that precise constraints for the behaviour of the bulk
cosmological constant, as well as the strict condition $E^{TT}_{\mu\nu} = 0$,
are needed for a Machian braneworld universe. Such constraints could be
intimately related to the initial conditions at the infinite past
surface of the brane observer.
We argue that if  the induced Einstein's equations on the
brane are reinterpreted as integral equations, it is possible to
arrive at conditions that are not evident in the usual differential
field equations, in special, the surface term allows for a much richer
interpretation of the interplay between MP, braneworlds, and the
cosmological constant. Such developments are presently being 
explored.

\subsection{Acknowledgements}

This work was supported by FAPESP under grant 01/06402-6 and
partially under grant 01/08310-1. Most part of this paper
was written during my post-doctoral activities at
Instituto Nacional de Pesquisas Espaciais (INPE),
Divis\~ao de Astrof\'{\i}sica (DAS/CEA),
Av. dos Astronautas, 1758, S. J. dos Campos, SP 12227-010, Brazil.

\end{document}